\documentclass[fleqn,usenatbib]{mnras}

\usepackage{newtxtext,newtxmath}

\usepackage[T1]{fontenc}

\DeclareRobustCommand{\VAN}[3]{#2}
\let\VANthebibliography\thebibliography
\def\thebibliography{\DeclareRobustCommand{\VAN}[3]{##3}\VANthebibliography}

\usepackage{booktabs}

\usepackage{graphicx}	
\usepackage{amsmath}	
\usepackage{orcidlink}

\title[BHXRB jet speeds]{Kinematics show consistency between stellar mass and supermassive black hole parent population jet speeds}

\author[Lilje et al.]{
Clara Lilje$^{\orcidlink{0009-0004-1555-5356}}$,$^{1}$\thanks{E-mail: Clara.Lilje@physics.ox.ac.uk}
Rob Fender$^{\orcidlink{0000-0002-3493-7737}}$,$^{1,2}$
and James H. Matthews$^{\orcidlink{0000-0002-5654-2744}}$,$^{1}$
\\
$^{1}$Department of Physics, Astrophysics, University of Oxford, Denys Wilkinson Building, Keble Road, Oxford, OX1 3RH, UK.\\
$^{2}$Department of Astronomy, University of Cape Town, Private Bag X3, Rondebosch, 7701, South Africa.
}

\date{\today}

\pubyear{\the\year{}}

\begin{document}
\label{firstpage}
\pagerange{\pageref{firstpage}--\pageref{lastpage}}
\maketitle

\begin{abstract}
Jets from stellar-mass and supermassive black holes provide the unique opportunity to study similar processes in two very different mass regimes. Historically, the apparent speeds of black hole x-ray binary (BHXRBs) jets have been observed to be lower than jet speeds from active galactic nuclei (AGN) and specifically blazars.
In this work, we show that selection effects could be the primary cause of the observed population differences. For the first time, it is possible to perform a statistical analysis of the underlying BHXRB jet Lorentz factor distribution. We use both the Anderson-Darling test and apply nested sampling to this problem. With Bayes factors, we confirm that the Lorentz factor distribution of BHXRBs is best described with a power law, the same model that has been applied to AGN jets. For a Lorentz factor distribution following $\rm N(\Gamma) \propto \Gamma^b$ we find a value for the exponent of $b=-2.64_{-0.55}^{+0.46}$. This exponent is consistent with values found in AGN population studies, within $1\sigma$ for \textit{Swift}-BAT and \textit{Fermi}-LAT selected AGN. The best-fit exponent for the radio selected MOJAVE sample is just above our $2 \sigma$ limit. This is a remarkable agreement given the different scales at which the jets are observed. The observed slower apparent speeds in BHXRBs are largely due to the much larger inclinations in this sample. Furthermore, nested sampling confirms that $\Gamma_{\rm max}$ is completely unconstrained using this method. Therefore, based on kinematics alone, BHXRB jets are broadly consistent with being just as relativistic as those from supermassive black holes.
\end{abstract}




\section{Introduction}
Astrophysical jets are produced by many accreting systems in our Universe, such as black holes (BHs), neutron stars (e.g. \citealt{munoz-darias_black_2014}) and white dwarfs (e.g. \citealt{sokoloski_uncovering_2008}) 
Generated from some of the most compact objects in our Universe, these jets cover some of the largest size scales, some reaching into the cosmic void \citep{oei_black_2024}. Therefore, improving our understanding of jets can enable a better understanding of the Universe at large as well as offering insight into small-scale accretion physics. 

Among compact objects, BHs are particularly intriguing laboratories for understanding jet formation across a large mass and distance scale range. BHs are some of the most uniquely scalable objects in the universe, as many of their observables are scaled only by their mass (e.g. \citealt{fender_eight_2007}). Properties such as spin or environment may introduce some scatter into the scaling relations, but the so-called fundamental plane of black hole activity shows remarkably consistent scaling of BH observables over eight orders of magnitude in BH mass \citep{merloni_fundamental_2003,falcke_scheme_2004,plotkin_using_2012}. 

In our own Galaxy, stellar-mass BHs produce jets in "X-ray binary" (XRB) systems, where they accrete mass from a stellar companion (e.g. \citealt{shakura_black_1973, meurs_number_1989,narayan_advection-dominated_1995,remillard_x-ray_2006}). These BHs usually have masses on the order of $M_{\rm BH}\sim 10M_{\odot}$ and extensive studies have investigated their variable behaviour (e.g. \citealt{casares_mass_2014,corral-santana_blackcat_2016,ingram_review_2020}). XRBs have been observed to undergo extreme spectral changes, which have been connected to changes of accretion disk - jet coupling throughout their duty cycle \citep{fender_towards_2004}. The ``hard state'' of XRBs is associated with a steady jet, while the transition from the ``soft'' to ``hard'' state is associated with radio flaring and the emission of large-scale transient jets (e.g. \citealt{bright_extremely_2020}). XRB systems that instead host neutron stars also produce relativistic jets, with similar, but not identical phenomenology (e.g. \citealt{munoz-darias_black_2014}).

At the other end of the mass scale there have been extensive studies of all classes of jet-producing supermassive black holes (SMBHs). The active, accreting BHs in some galaxies produce jets that can be much larger than their host galaxies \citep{blandford_relativistic_2019}. Although these objects have extremely long life times and we only ever see very small snapshots of their duty cycle, some show remarkable changing-state behavior (e.g. \citealt{meyer_late-time_2025}). There have been promising attempts to unify the different classes of Active galactic nuclei (AGN) into a framework of accretion states similar to XRBs (e.g. \citealt{kording_jet-dominated_2006,kording_accretion_2006,svoboda_agn_2017,fernandez-ontiveros_x-ray_2021,moravec_radio_2022}). 

A particularly suited sub-class for the study of jets are blazars - SMBHs with a relativistic jet closely aligned to our line of sight. As such these objects are affected by strong relativistic beaming and become extremely bright; therefore, they dominate flux-limited AGN samples at high energies and in the radio wavelengths. Resolving the kinematics of moving jet components in the brightest and most beamed AGN sources enables us to measure the speed of the jet and determine its Lorentz factor $\Gamma = \left( 1- \frac{v^2}{c^2} \right)^{-\frac{1}{2}}$ where $v$ is the speed of the jet and $c$ is the speed of light.

Historically, measured apparent XRB jet speeds have been found to be much lower than those of blazars. The first galactic superluminal source, GRS 1915+105 was measured to have mildly relativistic jets with $\Gamma \approx 2.5$ \citep{mirabel_superluminal_1994}, but this estimate was revised due to the impact of distance uncertainty on the measurement of Lorentz factor \citep{fender_uses_2003}. However, through dedicated observational campaigns, there is now a much larger sample of observed BHXRBs with large-scale transient jets and corresponding measured apparent speeds \citep{fender_speeds_2025,matthews_blast_2025}. Notably, the ThunderKAT program has played an instrumental part in revolutionizing our understanding of these jets \citep{fender_thunderkat_2016}. In addition, the recently discovered jets of 4U 1543-47 presented by \cite{zhang_jets_2025} show that an XRB can show high apparent speeds, with $\beta_{\rm app} \sim 5.3$ and hence $\Gamma > 5$. This suggests that the XRB jet speed sample is still largely limited by inclination and sample size.

In contrast, AGN and especially blazars have been observed with apparent speeds of up to $\beta_{\rm app} \sim 50$, much larger than all observed values for XRBs.
Larger number statistics of AGN make it possible to perform population modelling on fully flux-limited samples. These studies suggest that the underlying Lorentz factor distribution for AGN is well described by a power law \citep{lister_statistical_1997} with a mean Lorentz factors of $\Gamma \approx 10$ \citep{ajello_luminosity_2012,marcotulli_bass_2022}. As XRBs have been observed to only have apparent speeds of order a few, this could suggest an intrinsic difference in Lorentz factor distribution. XRBs however, are subject to very different selection effects, as they are detected in the X-ray, where the light from the infalling material dominates \citep{shakura_black_1973}. X-ray triggers of flaring XRBs enable us to obtain a sample of all local sources ejecting large-scale transient jets. As the sample is X-ray selected we anticipate an unbiased distribution of jet inclination angles. It is not possible to construct such a quasi-volume-limited, unbiased inclination sample for AGN, as their distance distribution is much more extended. Instead, flux-limited samples are constructed, which implicitly select for blazars. Due to their alignment with the line-of-sight they become boosted to be among the most bright and variable objects in the sky, but they represent a highly biased sample (e.g. \citealt{hovatta_relativistic_2020}).

The samples of BHs at both ends of the mass range provide a unique opportunity to understand jet and accretion physics, as well as providing insights into the BHs and their environments. In this work, we present a novel analysis of a sample of BHXRB jets and compare them to studies of the population of AGN jets. We take great care to disentangle selection biases and inclination effects to draw conclusions about the underlying physical mechanisms.
In Section \ref{sec:methods} we present the sample used in this paper and proceed to describe the parent population modelling procedure in Section \ref{sec:popmodel}. Then we constrain the model parameters with Anderson-Darling test in Section \ref{sec:adtest} and nested sampling using a Bayesian likelihood in Section \ref{sec:nestedSamp}.
In Section \ref{sec:Results} we present the results from these methods and discuss how they select a preferred Lorentz factor population model.
During the discussion in Section \ref{sec:Discuss} we compare our results to AGN population studies and show that there is a remarkable overlap in Lorentz factor distribution exponent parameter space between the two BH populations, which suggests that there may be similar jet physics across BH mass and size scales.
\section{Methods}
\label{sec:methods}
\subsection{XRB sample}
\label{sec:xrb_data}
\begin{figure*}
    \centering
    \includegraphics[width=0.8\textwidth]{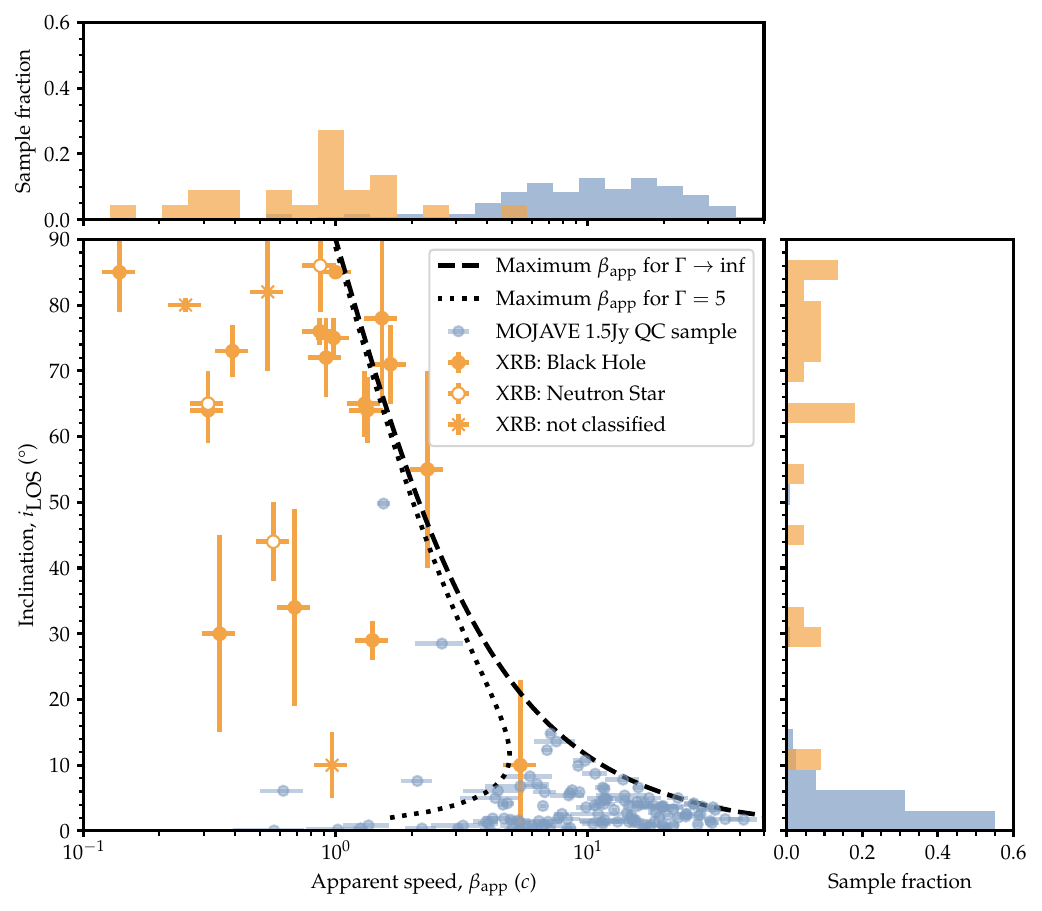}
    \caption{Comparison of the inclination angle and apparent speed distributions of XRB and AGN jets. X-ray binary sample of apparent speed and inclinations used in this work as presented in \protect\cite{fender_speeds_2025} shown as orange data points. The marker symbol indicates the class of object. The blue data points show the MOJAVE 1.5 Jy Quarter Century sample \citep{lister_mojave_2009,homan_mojave_2021}. The dotted black curve shows the maximum observable apparent speed due to inclination effects at an infinitely high Lorentz factor and at $\Gamma=5$. The histograms are marginalized over each axis and show the sample fraction of each of the populations. The populations occupy very different regions of parameter space.}
    \label{fig:xrb_sample}
\end{figure*}
The XRB data used in the analysis of jet speeds in this work is presented by \cite{fender_speeds_2025}. It presents the first statistically significant sample of directly measured speeds of jets launched from XRBs. Inclination and distance measurements for the sources are also compiled in their work.

The apparent speed of each source is estimated from the observed proper motion of the spatially resolved ejecta traveling away from the XRB core (location of the BH), usually after the transition from the hard to the soft state. In particular, for known distance to the object, $d ~[\text{kpc}]$, and observed proper motion, $\mu ~[\text{mas d}^{-1}]$, the apparent speed, $\beta_{\rm app}$, as a fraction of $c$ is given by 
\begin{equation}
\label{eq:propmotion}
    \beta_{\rm app}=\frac{\mu d}{173 \text{[kpc  mas d}^{-1}]} .
\end{equation}
Some sources have shown both highly- and mildly-relativistic speeds, which have been interpreted to be two different jets: one that is spin-locked to the black hole and one that is launched from further out in the disk \citep{fender_speeds_2025}.

In Figure \ref{fig:xrb_sample} we present this XRB sample with orange data points in inclination versus apparent speed space, alongside the flux-limited MOJAVE $1.5$ Jy Quarter Century sample of AGN in the blue points.
In the XRB sample, there is significant uncertainty in the inclination measurement. In Figure \ref{fig:xrb_sample} the inclination error bars represent a range of possible values given in the literature. Therefore, they should be interpreted as very approximate uniform error distributions. 
Even when accounting for these large uncertainties in inclination, the overall sample distribution is consistent with an isotropic angle distribution when performing the Kolmogorov-Smirnov (KS) test, giving a p-value of $p=0.5878$. 
The sample is indeed expected to be approximately isotropic in inclination in 3D space, as the sources are discovered due to their x-ray emission. As this radiation escapes approximately isotropically from the accreting material, we would not expect this selection mechanism to select for any specific inclinations towards the line-of-sight \citep{shakura_black_1973}.

The sample of XRBs consists of $15$ BH systems with measured apparent speeds. One source, MAXI J1820$+$070, shows two different jet speeds, inclined on the same axis to the line of sight \citep{wood_varying_2021}. There are $2$ unknown accretors in the sample. One of them, Cygnus X-3, also shows two different jet speeds, which are at two completely different inclinations \citep{schalinski_vlbi_1995, mioduszewski_one-sided_2001}. Lastly, there are $3$ Neutron stars in this sample. They show less relativistic speeds than the BH binaries (\citealt{fender_speeds_2025}; see also \citealt{russell_thermonuclear_2024}). We assume apparent speed uncertainties of $15\%$ in the figure to approximately account for distance and proper motion uncertainties. There is a wide range of confidence in the measured literature values of apparent speeds \citep{fender_speeds_2025}. Many proper motion measurements are quite precise and constrained with gaussian errors, which are usually of order $\sim 10\%$ (e.g. \citealt{corbel_discovery_2005,rushton_resolved_2017,russell_disk-jet_2019,wood_time-dependent_2023}). There are some exceptions where uncertainties are much smaller \citep{espinasse_relativistic_2020} or given with uniform ranges (e.g. \citealt{hjellming_episodic_1995,chauhan_broadband_2021}). The contribution to the apparent speed uncertainty from distance measurements are usually larger. There are diverse methods of measuring distance and some give reliable gaussian errors, which are of $\sim 5-10\%$ (e.g. \citealt{miller-jones_first_2009,shaposhnikov_discovery_2010,orosz_improved_2011}). Some sources only have best guess distances within uniform ranges, which range around $\sim 20\%$ (e.g. \citealt{tingay_relativistic_1995,zdziarski_x-ray_2019, wood_time-dependent_2023}). The "effective $1\sigma$" gaussian errors of a $20\%$ uniform range are $\sim 11.5 \%$ and propagating these forward with the conservative $\sim 10\%$ gaussian proper motion uncertainties results in an average $\sim 15\%$ uncertainty on $\beta_{\rm app}$. We only choose this representative $\beta_{\rm app}$ uncertainty to plot. We separately account for the impact of the large uniform distance uncertainties on our results by resampling as described in section \ref{sec:popmodel}.

\begin{figure}
    \centering
    \includegraphics[width=1\linewidth]{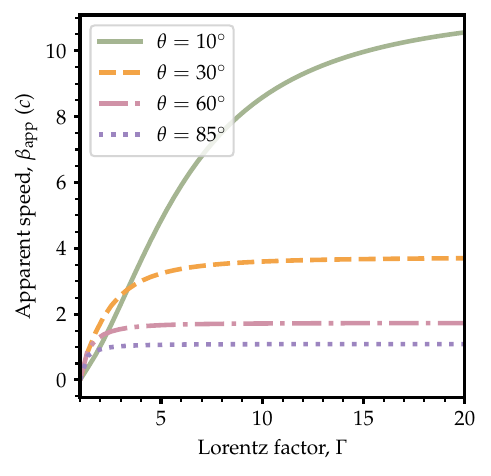}
    \caption{The change in observed apparent speed with changing intrinsic Lorentz factors at different inclination angles. This figure shows the asymptotic flattening of the apparent speed at high inclinations. This effect affects the XRB sample strongly.}
    \label{fig:lorentz-beta}
\end{figure}

Figure \ref{fig:xrb_sample} also demonstrates that there is a maximum observable apparent speed $\beta_{\rm app}$ at each inclination, no matter the underlying Lorentz factor. This is due to the infinite Lorentz factor limit of the equation of apparent superluminal motion, also demonstrated in Figure \ref{fig:lorentz-beta} where we show the apparent speed $\beta_{\rm app}$ calculated as 
\begin{equation}
    \label{eqn:apparentspeed}
    \beta_{\rm app} = \frac{(1-\Gamma^{-2})^{\frac{1}{2}} \sin{\theta}}{1-(1-\Gamma^{-2})^{\frac{1}{2}}\cos{\theta}} \, .
\end{equation}
Figure \ref{fig:lorentz-beta} shows that, at high inclinations, the observable apparent speed flattens asymptotically regardless of the underlying Lorentz factor. This can also be obtained mathematically, if we take the limit of $\Gamma \rightarrow \infty$, such that
\begin{equation}
    \label{eqn:limitapparentspeed}
    \beta_{\rm app} = \frac{\sin{\theta}}{1-\cos{\theta}} \, .
\end{equation}
which is valid for $0 \leq \theta \leq \frac{\pi}{2} $
This asymptotic value is plotted as the dashed line in Figure \ref{fig:xrb_sample}. The XRB sample largely adheres to this upper limit when uncertainties are taken into account. A direct consequence of most XRBs being off-axis is that even if there was a large change in $\Gamma$ for any of the high-inclination sources, this would only lead to a subtle change in $\beta_{\rm app}$ and therefore impact the observed apparent speed distribution only marginally.

We also plot the AGN in the MOJAVE 1.5 Jy Quarter Century sample \citep{lister_mojave_2009} in Fig. \ref{fig:xrb_sample}, which is also described in Section \ref{sec:blazarstudies}. These sources occupy a different region in Figure \ref{fig:xrb_sample} when compared to the XRBs. Most sources in this sample have inclination angles of less than $10^\circ$ with a peak at $\sim 1-2^\circ$, which classifies them as blazars \citep{urry_unified_1995}. Their apparent speed distribution peaks at $\beta_{\rm app}\sim10$. They also follow the black dashed line of asymptotically limited maximum observable apparent speeds.
The AGN sample is dominated by these low inclination - high $\Gamma$ sources as it is flux-limited. This selects for blazars as they appear brighter as a result of Doppler boosting. The sample is not selected with direct constraints on the inclination angle; these appear implicitly through the flux requirements.
Figure \ref{fig:xrb_sample} demonstrates the selection biases that affect these two populations well. In particular, note that the marginalised histograms show that there is very little overlap in this parameter space between the two populations due to their selection biases. The very different inclination angle distributions leads naturally to very different observed speed distributions.

Since we are aiming to compare BHs across the mass range, we exclude neutron stars from the analysis. We keep both unknown sources in the sample, as there is evidence that both could be powered by a BH (SS433: \cite{bowler_ss_2018,cherepashchuk_discovery_2021}; Cyg X-3: \cite{zdziarski_cyg_2013}. The inclusion of these objects does not significantly affect the results of this work. The objects in this reduced sample will be referred to as BHXRBs (black hole x-ray binaries) from this point onward.
\subsection{Parent Population modelling}
\label{sec:popmodel}
To investigate the intrinsic Lorentz factor distribution of BHXRBs, we employ Monte Carlo parent population modelling. Similar analyses have been conducted for flux-limited AGN samples, which are significantly larger (on the order of 100 sources compared to 19 BHXRB jets) \citep{ajello_luminosity_2012,liodakis_population_2015,lister_mojave_2019,marcotulli_bass_2022}. However, AGN populations are subject to more complex selection biases, including flux limits, redshift distribution non-uniformity, and strong Doppler beaming, which preferentially selects blazar sources aligned closely with the line of sight.

To model the BHXRB population, we assume an underlying isotropic angle distribution in 3D space. As discussed in section \ref{sec:xrb_data}, this is consistent with the data. To sample a variable from a given probability distribution, we can sample the inverse of the cumulative density function (CDF) of the variable. For the (unbiased) BHXRB inclination angle distribution, the probability density function (PDF) is 
\begin{equation}
\label{eqn:pdfangle}
    p(\theta) = \sin\theta \, ,
\end{equation}
for $0\leq \theta \leq \pi/2$ with $\theta$ expressed in radians, since we see the 2D projection of the angle in the sky. The CDF is trivially
\begin{equation}
\label{eqn:cdfangle}
    \rm CDF(\theta) = 1 - \cos{\theta} \, .
\end{equation}
To obtain a value of $\theta$ we sample a uniform random variable $u = U(0,1)$ and solve
\begin{equation}
\label{eqn:inversecdfangle}
    \theta(u) = \arccos(u)
\end{equation}
for $0\leq u \leq 1$ where we have made the transformation $u' = 1+u$ to symmetry arguments.
The intrinsic distribution for Lorentz factors is also sampled with the same method. A power law is the description of choice for this process in the AGN literature \citep{lister_statistical_1997}. We justify this choice for BHXRBs in Section \ref{sec:modelsel}.
The probability density function (PDF) of a power law Lorentz factor distribution is:
\begin{equation}
    \rm N(\Gamma) = \frac{b+1}{\Gamma_{\rm max}^{b+1}-\Gamma_{\rm min}^{b+1}}\Gamma^b
\end{equation}
for $\Gamma \in [\Gamma_{\rm min},\Gamma_{\rm max}]$ and $0$ otherwise.
The CDF is trivially
\begin{equation}
    \mathrm{CDF}(\Gamma) = 
        \dfrac{\Gamma^{b+1} - \Gamma_{\rm min}^{b+1}}{\Gamma_{\rm max}^{b+1} - \Gamma_{\rm min}^{b+1}} \, ,
\end{equation}
in the range $\Gamma \in [\Gamma_{\rm min},\Gamma_{\rm max}]$
and for the case $b=-1$:
\begin{equation}
    \mathrm{CDF}(\Gamma) = 
        \dfrac{\ln\left( \Gamma / \Gamma_{\rm min} \right)}{\ln\left( \Gamma_{\rm max} / \Gamma_{\rm min} \right)} \, ,
\end{equation}
This can be sampled with a uniform random variable $u = U(0,1)$ and inverted to give (for $b \neq -1)$:
\begin{equation}
\label{eqn:powerlawinversecdf}
  \Gamma(u) = \left(u(\Gamma_{\text{max}}^{b+1}-\Gamma_{\text{min}}^{b+1}) + \Gamma_{\text{min}}^{b+1}\right)^{\frac{1}{b+1}}  \, ,
\end{equation}
for $\Gamma \in [\Gamma_{\rm min},\Gamma_{\rm max}]$.
For the special case at $b=-1$ we obtain instead:
\begin{equation}
\label{eqn:specialcase}
  \Gamma(u) = \Gamma_{\text{min}} \exp \left[(\ln{\Gamma_{\text{max}}}-\ln{\Gamma_{\text{min}}})u\right] \, .
\end{equation}
\subsubsection{Simulation procedure}
\label{sec:simprod}
Once the parent population can be fully described with the assumed probability distributions, we use the following method to generate a Monte Carlo parent distribution for BHXRB jets:
\begin{enumerate}
    \item Choose the parameter values $b$, $\Gamma_{\text{min}}$ and $\Gamma_{\text{max}}$ for the parent distributions.
    \item Draw a value of inclination angle $\theta$ using equation \ref{eqn:inversecdfangle}. 
    \item Determine an intrinsic Lorentz factor $\Gamma$ for each jet in the parent sample using equation \ref{eqn:powerlawinversecdf} or \ref{eqn:specialcase}.
    \item Calculate the apparent speed, $\beta_{\rm app}$, of each source following equation \ref{eqn:apparentspeed}.
\end{enumerate}
Once the distribution of apparent speeds is built up, we compare to the observed distribution of apparent speed using both the Anderson-Darling test (sec \ref{sec:adtest}) and an approximate Bayesian likelihood (sec \ref{sec:nestedSamp}).
As the distances of our sample are accounted for in the calculation of the apparent speeds of the data (see equation \ref{eq:propmotion}) we do not need to include a distance distribution in our modelling. Since there is, however, a significant uncertainty associated with the distance measurement and this directly affects the observed apparent speed distribution, we resample the distances with uniform uncertainties up to $40 \%$ and recompute the apparent speeds of our data set. We find the mean $b$ value to remain consistent within $1 \sigma$. This is largely due to the broad range covered by the posterior and is likely to change once the sample size grows. 

A significant advantage of this methodology is that we can directly test the apparent speeds of the sources, which is a much more observationally robust quantity. Studies such as \cite{fender_speeds_2025} rely on inferred Lorentz factors, which are often only lower limits due to the asymptotic flattening of the apparent speeds as demonstrated in Fig. \ref{fig:lorentz-beta}. Therefore, directly testing the apparent speeds circumvents any potential issues with accounting for lower limits. 

\subsection{Anderson-Darling test}
\label{sec:adtest}
A common way to determine whether two distributions are drawn from the same distribution is the Anderson-Darling (AD) test \citep{anderson_asymptotic_1952}. We use this test statistic to determine which simulated parent population best matches the observed sample of BHXRBs. We followed this procedure:
\begin{enumerate}
    \item Fix values of $\Gamma_{\text{min} }=1,\Gamma_{\text{max}}=50$.
    \item Step through defined prior range of values for exponent $b$ uniformly.
    \item Generate apparent speed sample of 19 sources from chosen distributions as described in section \ref{sec:popmodel}.
    \item Perform AD test between observed apparent speeds from \cite{fender_speeds_2025} and simulated sample using \texttt{scipy.stats.anderson\_ksamp} and a permutation test for exact p-values \citep{scholz_k-sample_1987}.
    \item Record p-value of test
    \item Repeat steps (ii) to (iv) 100 times for each value of $b$.
    \item Record fraction of trials in which the null hypothesis has been rejected at the chosen confidence interval.
\end{enumerate}
We perform two separate analyses with thresholds of $p<0.05$ and $p<0.01$ to obtain an approximate confidence interval estimate. When the null hypothesis is rejected with the AD test it means that the samples are likely not from the same parent distribution. Therefore, a low rejection fraction means that most of the synthetic distributions are consistent with the data, which is indicative of - although not formally the same as - a ``well-fit'' distribution.

We tested varying both Lorentz factor limits $\Gamma_{\text{min}}$ and $\Gamma_{\text{max}}$, but the choice of $\Gamma_{\text{min}}$ is physically motivated, as the Lorentz factor cannot be smaller than $1$. As indicated by Figure \ref{fig:lorentz-beta}, even large changes in the maximum $\Gamma$ lead to small changes in the observed $\beta_{\rm app}$ distribution when most of the sample is observed at high inclinations. As the observed apparent speed is a lower limit to the Lorentz factor, the $\Gamma_{\text{max}}$ must be larger than the maximum observed apparent speed, but is quite unbounded towards higher values. Changing $\Gamma_{\text{max}}$ does not affect the distribution significantly.
The parameter space for $b$ was informed by analysis of AGN populations, where the power law exponent is found to be in the region of $b \in [-5,-0.5]$. Additionally, a coarse search over the range $b \in [-10,0]$ indicated a preferred parameter space for BHXRBs around $b \in [-6,0]$; this range was then explored with a finer grid of $\Delta b \simeq 0.1$.

While this analysis offers insight into the regions of parameter space where the rejection fraction is minimized — indicating which values of $b$ are most consistent with the observed distribution — it does not constitute complete Bayesian inference and therefore does not yield a posterior distribution or credible intervals.
\subsection{Nested Sampling}
\label{sec:nestedSamp}
To obtain posteriors for the parameters and extend our analysis to also include constraints on $\Gamma_{\text{min}}$ and $\Gamma_{\text{max}}$ we use nested sampling to explore our parameter space \citep{skilling_nested_2004,skilling_nested_2006}. This is implemented using the python package \texttt{dynesty} \citep{speagle_dynesty_2020,koposov_joshspeagledynesty_2024}. As shown in Table \ref{tab:priors}, we are using uniform priors for all parameters, which are non-informative. This is appropriate, as there is no physical intuition as to why a particular value should be preferred. The bounds of the parameters are physically motivated. It is not expected that the Lorentz factor PDF increases with increasing $\Gamma$, as the sample contains more mildly relativistic sources \cite{fender_speeds_2025} even below the minimum observable apparent speed cut-off. Therefore, positive values of $b$ are excluded. A uniform underlying Lorentz factor distribution is plausible and $b=0$ is therefore included in the priors. Additionally, the bounds for $b$ are informed by the previous analysis using the AD test, which show a minimum around values of $b\sim-3$.

It is evident from Figure \ref{fig:lorentz-beta} that the apparent speed is a lower limit to the Lorentz factor. Therefore, it is required that our distribution extends at least until the fastest measured apparent speed, which implies $\Gamma_{\text{max}}>5$. There is no physical reason to place a tight constraint on the upper end of this prior and since we are also aiming to compare our sample to AGN we wish to explore a broad range of $\Gamma_{\text{max}}$.
The upper bound of $\Gamma_{\text{min}}$ is set by observations of mildly relativistic jets in BHXRBs as described in \cite{fender_speeds_2025}.
The breadth of the priors has been tested; while very narrow priors can affect results, the chosen broad priors yields a converged solution.
\begin{table}
    \centering
    \begin{tabular}{|c|c|c|c|}
    \toprule
    \toprule
        & Lower bound & Upper bound & Type \\ \midrule
        $b$ & $-6.0$ & $0.0$ & Uniform \\ 
        $\Gamma_{\text{min}}$ & $1.0$ & $3.0$ & Uniform \\ 
        $\Gamma_{\text{max}}$ & $5.0$ & $200.0$ & Uniform \\
        \bottomrule
    \end{tabular}
    \caption{Prior parameters used in the nested sampling analysis for the Lorentz factor parent population distribution.}
    \label{tab:priors}
\end{table}
\subsubsection{Likelihood choice}

The nested sampling algorithm has to be informed by a logarithmic likelihood.
Due to the sparseness of the data, we choose to use a kernel density estimator (KDE) \citep{rosenblatt_remarks_1956,parzen_estimation_1962} to approximate the likelihood.
This approach utilizes the fact that the simulated distribution can be well described by a unimodal distribution, examples are shown in Figure \ref{fig:ap_model_comp}. A unimodal distribution is well described by a Gaussian KDE with an appropriate choice of band width $h$ described by 
\begin{equation}
\label{eqn:kde}
  P_{model}(x) = \frac{1}{nh}\sum_{i =1}^{n} K\left(\frac{x - x_i}{h}\right)  
\end{equation}
and 
\begin{equation}
\label{eqn:bandwidth}
    K(x) = \frac{1}{\sqrt{2 \pi}}\exp\left( -\frac{x^2}{2}\right)
\end{equation}
A KDE is a nonparametric way to estimate the probability density function of a distribution, which is especially suited to our application, as we do not have an analytical function to describe our model distribution due to the convolution of inclination and intrinsic speed effects.
The bandwidth optimisation is automatically determined within \texttt{scipy.stats.gaussian\_kde} \citep{bashtannyk_bandwidth_2001,heidenreich_bandwidth_2013,scott_multivariate_2015} and the results converge when sampling an appropriate amount of model sources. These convergence tests informed the choice of $\rm N_{\rm sim}$. Examples of the model KDEs can be found in Figure \ref{fig:ap_model_comp}.

Since the KDE gives the probability of each point, we define our log-likelihood as
\begin{equation}
\label{eqn:likelihood}
    \log(L) = \sum_i \log (P_{\text{model}}(\beta_{\text{obs,}i})).
\end{equation}
This is a likelihood that corresponds to the discretized version of the Gaussian likelihood: 
\begin{equation}
\label{eqn:gaussianlikelihood}
    P(D|M) = \int P(D|\theta,M) P(\theta|M)d\theta
\end{equation}
As the Bayesian likelihood is defined as the probability of the data given the model, the KDE returns exactly this when evaluated at the observed data points. Similar approaches have been used in other cases in the literature when tuning KDE hyperparameters \citep{jones_maximum_2009,shang_bayesian_2020}, and KDE likelihoods have been used in the medical field to model airways \citep{rudyanto_modeling_2013}.

The nested sampling is then implemented as follows:
\begin{enumerate}
    \item Obtain parameter values from the walk through parameter space performed by the nested sampling algorithm implemented in \texttt{dynesty}. This is informed by the prior distribution shown in Table \ref{tab:priors}.
    \item Generate apparent speed sample of $\rm N_{\rm sim} = 10^5$ sources from chosen distribution as described in section \ref{sec:simprod}.
    \item Determine a KDE of the simulated data sample, $P_{\text{model}}(\beta_{\rm app})$, following equation \ref{eqn:kde}, implemented in \texttt{scipy.stats.gaussian\_kde}.
    \item Evaluate the KDE at the observed apparent speeds to obtain $P_{\text{model}}(\beta_{\text{obs,}i})$
    \item Obtain log-likelihood from equation \ref{eqn:likelihood} as described above.
\end{enumerate}

The nested sampling algorithm then iterates over parameter space while maximizing the overall likelihood. In addition, \texttt{dynesty} calculates the evidence automatically for each point in parameter space. Therefore, we also obtain the importance weights for each point, which are then used to obtain posterior distributions for each parameter \citep{skilling_nested_2006}. These are shown in Figure \ref{fig:NS_corner}.

\subsubsection{Bayes factor model comparison}
\label{sec:bfModel}
While the power law model for Lorentz factors is a reasonable choice given the literature on AGN, nested sampling enables us to use evidence to compare different Lorentz factor distribution models using Bayes factors.
The Bayes factor is the ratio of two marginal likelihoods and indicates which model can be preferred over another \citep{kass_bayes_1995}.
The Bayes factor $K$ is given by:
\begin{equation}
\label{eqn:bayesfactor}
    K = \frac{\text{Pr}(D|M_1)}{\text{Pr}(D|M_2)} = \frac{\frac{\text{Pr}(D|M_1)\text{Pr}(D)}{\text{Pr}(M_1)}}{\frac{\text{Pr}(D|M_2)\text{Pr}(D)}{\text{Pr}(M_2)}} = \frac{\text{Pr}(D|M_1)\text{Pr}(M_2)}{\text{Pr}(D|M_2)\text{Pr}(M_1)}
\end{equation}

We assume no prior knowledge as to which model should be preferred ($\text{Pr}(M_1) = \text{Pr}(M_2) =1$), so the Bayes factor is the ratio of the posterior probabilities, which is equivalent to the evidence, $Z$. We optimise the parameters for each model using nested sampling as described above and compute the Bayes factor as
\begin{equation}
    \log(K) =\log(Z_1) - \log(Z_2) .
\end{equation}
The other Lorentz factor models tested are a truncated Gaussian, a truncated exponential and Gamma function. The formulae are given in the Appendix \ref{app:altLorentz}. 

\section{Results}
\label{sec:Results}
\begin{figure}
    \centering
    \includegraphics[width=1.0\linewidth]{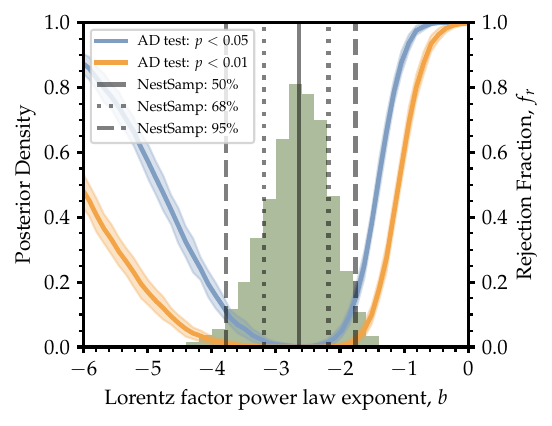}
    \caption{Comparison between Nested Sampling approach as well as Anderson-Darling test statistic rejection fraction for the exponent of the Lorentz factor power law describing the distribution in BHXRBs. The green histogram is the nested sampling posterior, with the posterior density shown on the left y-axis. The corresponding quantiles are shown with the gray lines. The right y-axis shows the rejection fraction, where the blue line is the rejection fraction at $p<0.05$, which corresponds to a confidence interval of $95\%$. And the orange line at $p<0.01$ which corresponds to a confidence interval of $99\%$ when using the Anderson-Darling test. }
    \label{fig:nestedsampling-versus-AD}
\end{figure}
We have presented two methods to determine the shape of the parent population Lorentz factor distribution for BHXRB jets. These enable us to constrain population-wide statistics on the jet speeds of these objects for the first time. Additionally, we are able to account for selection effects to compare this population to BHs at the other end of the mass range.
\subsection{Model Selection}
\label{sec:modelsel}

In Figure \ref{fig:nestedsampling-versus-AD} we show the constraints on $b$ obtained from the AD test, by sampling $b \in [-6.0,0.0]$ as the orange ($99\%$ confidence interval) and the blue curve ($95\%$ confidence interval). Both the rejection fractions at $p<0.05$ and at $p<0.01$ indicate that there is a preferred parameter space for $b$ around $b\approx-2.7$, but there is a large degeneracy of parameter values at which the rejection fraction is very low. When choosing the confidence interval of $95\%$ and finding the minimum and maximum value of $b$ for which the rejection fraction is $f_{r} <0.05$ we find that $b$ is bounded by $b \in [-3.47,-2.08]$ with a median of $b=-2.78$. 

As anticipated, since we are placing a higher threshold on determining that the modelled and observed distributions are different, the curve at the $99\%$ confidence level is shallower and exhibits a broader minimum. Both curves are observed to rise more steeply toward increasing values of $b$, indicating that this region of parameter space is more sharply constrained. This trend is consistent with theoretical expectations, as the asymptotic flattening of the apparent speeds caused by the higher inclinations in XRBs due to their unbiased angle distribution causes the power-law distribution to appear more uniform at low $b$. As a result, $b$ is less tightly constrained toward lower values.

\begin{figure}
    \centering
    \includegraphics[width=1.0\linewidth]{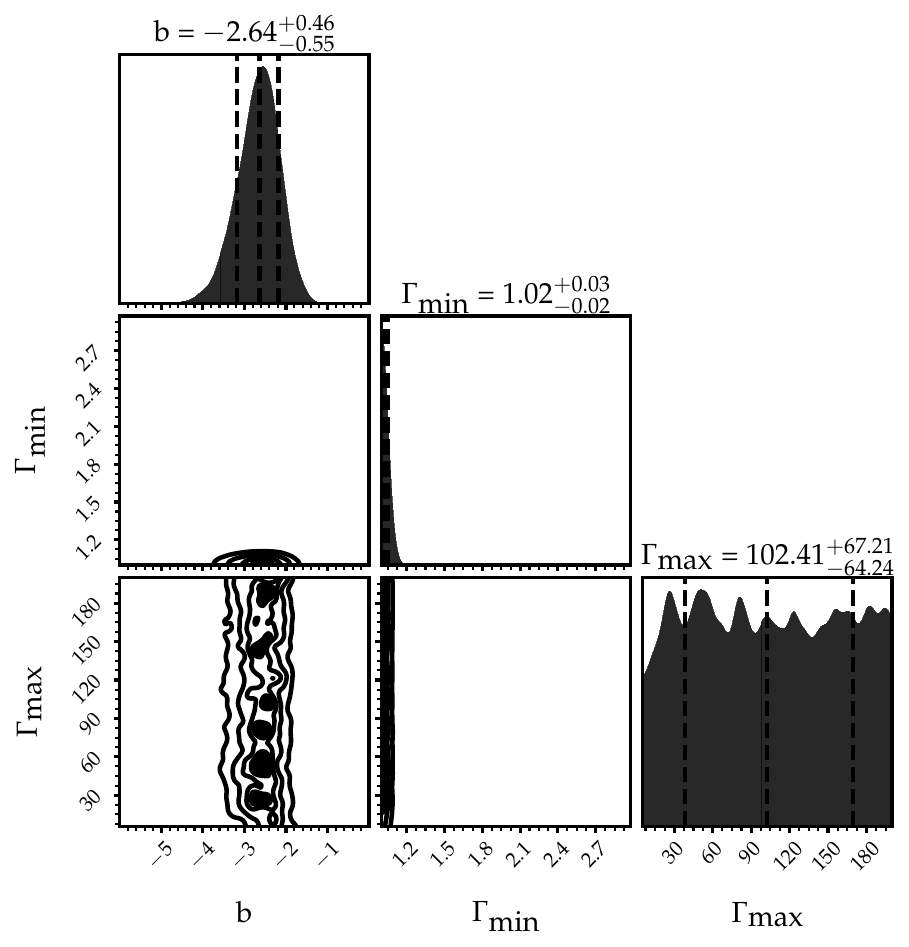}
    \caption{The results of the Nested Sampling analysis of the BHXRB sample. The plots on the diagonal show the marginalised 1D histograms for each parameter, while the other plots show 2D marginalised distributions. The quoted uncertainties are $1\sigma$ limits.}
    \label{fig:NS_corner}
\end{figure}

Tighter constraints and a better-defined posterior on all parameters are derived from nested sampling. This analysis returns marginalised posteriors for all three parameters as shown in Figure \ref{fig:NS_corner}. The uncertainties quoted are a $1\sigma$ interval.
The figure demonstrates that $b$ and $\Gamma_{\text{min}}$ are well constrained, while $\Gamma_{\text{max}}$ is not. As expected, the value of $\Gamma_{\text{min}}$ is consistent with $\Gamma=1$ (stationary). Any deviations from this limit would be caused by the limited observing time of these systems. Such as the case where it is not possible to resolve an ejection travelling at extremely low speeds away from the core in the observational time frame of decades.

The upper limit of the Lorentz factor, $\Gamma_{\text{max}}$, remains entirely unconstrained by this method. Although $\Gamma_{\text{max}}$ affects the overall normalization, its influence on the distribution shape is minimal, as the probability of extremely high Lorentz factor jets rapidly decreases under a power-law probability distribution function. As a result, due to plateau-ing of the apparent speeds and the small sample size, the contribution of high-$\Gamma$ jets to the observed population is small and difficult to constrain precisely. Any structure within the posterior in Fig. \ref{fig:NS_corner} is not likely to be real, but rather an artifact of the walk through parameter space performed by the nested sampling algorithm.

Lastly, this method provides interesting constraints on the parameter $b$ at a value of $b = -2.64_{-0.55}^{+0.46}$. The shape of this posterior is in very good agreement with the A-D analysis. The nested sampling posterior is also shown in Figure \ref{fig:nestedsampling-versus-AD} on the same axes as the result given by the AD rejection tests. The nested sampling result displays a similar shape with a flatter slope towards lower $b$ and a steeper slope towards $b=0$. The minimum of the AD rejection fractions and the maximum of the Nested Sampling also coincide very well. This validates the nested sampling as an appropriate method to apply to this problem and additionally indicates that the chosen likelihood performs well.  
As described in Section \ref{sec:bfModel} the Bayes factor can discriminate between multiple models for the Lorentz factor distribution. The Bayes factor is interpreted with the Jeffreys scale \citep{jeffreys_theory_1961} or with the table provided by \cite{kass_bayes_1995}. Any negative value of $\log(K)$ is evidence for Model 2 and any positive value indicates evidence for Model 1. The results are shown in Table \ref{tab:bayes_factors}, where Model 1 is displayed in the rows, and Model 2 in the columns of the table. The evidence points to the power law distribution being preferred over all other models. The evidence is classified as decisive for the power law in all cases. This justifies our choice of power law model given the prior information available to us. As can be seen in Figure \ref{fig:ap_model_comp} this decisive evidence is key since the best-fit observed apparent speed distribution is very similar for all tested underlying Lorentz factor distributions. The Bayes factor is insensitive to the amount of free parameters in a model. It should however be noted that there is evidence that broader priors influence the evidence  calculation when using nested sampling. Since uniform priors are used and many are similarly broad across models, this effect is negligible for this analysis. 

To constrain the effect of outlier sources, such as 4U 1543-47, we remove this highly relativistic, low inclination source and re-run the analysis. Without this source the median of the posterior shifts towards lower values $b \sim -3.1$, which remains within $1 \sigma$ of the result including the source. And the posterior space covered remains very similar, only extending further towards $b=-6$. The posterior for $\Gamma_{\rm max}$ is similarly unconstrained when 4U 1543-47 is removed.
\begin{table}
    \centering
    \begin{tabular}{c|c|c|c|c}
    \toprule
    \toprule
        $\log(K)$ & Power law & Exponential & Gaussian  & Gamma\\ \midrule
        Power law & N/A & $2.504$ & $8.427$ & $3.036$\\ 
        Exponential & $-2.504$ & N/A & $5.924$ & $0.532$\\ 
        Gaussian & $-8.427$& $-5.924$ & N/A & $-5.392$\\ 
        Gamma & $-3.036$ & $-0.532$ & $5.392$ & N/A\\ \bottomrule
    \end{tabular}
    \caption{Model 1 shown on the rows of the table and Model 2 shown in the columns. The logarithm of the Bayes factor $K$ given as a log value in the corresponding cells.}
    \label{tab:bayes_factors}
\end{table}
\section{Discussion}
\label{sec:Discuss}
The results presented show convincingly that the parent population of Lorentz factors for BHXRBs can be described by a power law. This analysis has never been done for these objects and has only recently become possible due to dedicated long-term follow-up to track transient jets from these sources. 

In addition, we have presented a new form of analysis for parent population modelling. Nested sampling presents many advantages over traditional methods, such as AD tests, as it provides full posteriors for all parameters. The choice of likelihood also allows for analysis on smaller datasets. 

A key result of the nested sampling analysis is that the upper limit of the Lorentz factor $\Gamma_{\text{max}}$ is completely unconstrained by this analysis. This is clearly the result of inclination effects on the shape of the observed distribution, since the maximum Lorentz factor can only be well-constrained by low $i$ and high $\Gamma$ sources. Here, the analysis is limited by the composition of the sample, as there is only one of these sources present. Work by \cite{savard_relativistic_2025} indicates that other methods will be able to constrain the maximum Lorentz factor of these ejections. Simulations show that once the internal energy of the ejected blobs is too high, they behave like Sedov blast waves and disrupt while propagating. \cite{miller-jones_opening_2006} constrain the Lorentz factors of large-scale jets of XRBs through their opening angles. This results in large $\Gamma > 10$, but this requirement could be relaxed if the jets are confined. Additional constraints on the maximum initial Lorentz factor of BHXRBs can be constrained by modelling the motion of the ejecta with blast-wave models \citep{carotenuto_constraining_2024} or joint radiative and kinematic modelling \citep{cooper_joint_2025}. These approaches find moderate Lorentz factors for BHXRBs.

The posteriors of the power law exponent $b$ provide an intriguing opportunity to compare BHXRBs to other jetted BHs at completely different mass scales. The nested sampling algorithm enables us to place good constraints on the shape of the parent Lorentz factor distribution.

\subsection{AGN population studies}
\label{sec:blazarstudies}
As previously discussed, AGN are an ideal comparison sample up to eight orders of magnitude above the BHXRBs in the mass scale. Due to their extreme selection effects, many of the population studies performed use complete flux-limited samples, such as \cite{ajello_luminosity_2012,liodakis_population_2015,lister_mojave_2019,marcotulli_bass_2022}. All of these analyses use a power law to fit the parent Lorentz factor distribution of blazars. The samples differ in the wavelength band they are selected in.

\subsubsection{MOJAVE 1.5 Jy Quarter Century sample}
\cite{lister_mojave_2019} perform population studies on the MOJAVE 1.5 Jy Quarter Century sample \citep{lister_mojave_2009}, which is plotted in Figure \ref{fig:xrb_sample}. This is a radio-selected sample of AGN that reach $1.5 $ Jy in Very Long Baseline Array (VLBA) correlated flux density at least once within the observing period of more than 20 years. The sources are observed with the VLBA to track the parsec-scale jets and the kinematics of individual jet components. \cite{lister_mojave_2019} use the AD test p-value to compare the simulated distributions to their observed sample of 174 radio-selected flat spectrum radio quasars (FSRQs). The authors fit the Lorentz factor distribution $\rm N(\Gamma) \propto \Gamma^b$ simultaneously with the beamed luminosity, the redshift and the flux distribution.
Although this allows for more of the available data to inform the best-fit model, there are more assumptions about the parent population distributions required. These are presented in \cite{lister_mojave_2019}. The resulting best-fit model with the highest AD p-value for all 4 distributions gives an exponent of $b=-1.4 \pm0.2$ for the $\Gamma$ distribution. The authors choose $\Gamma_{\text{min}}=1.25$ and $\Gamma_{\text{max}}=50$ as their distribution limits which are based on the maximum instantaneously measured speed of a jet feature. The lower limit is based on the relative prominence of radio cores by \cite{mullin_bayesian_2009}.

\subsubsection{BAT AGN Spectroscopic survey (BASS)}
Another large AGN sample is the sub-sample of the BAT AGN Spectroscopic survey (BASS) \citep{paliya_bat_2019}. This sample is selected in the hard x-ray with the \textit{Swift}-Burst Alert Telescope (BAT) in the 14-195 keV range and limited at a flux greater than $F_{14-195 \text{~keV}} > 5.4 \times10^{-12} \text{erg~cm}^{-2}\text{s}^{-1}$. In \cite{marcotulli_bass_2022} the parent population of the FSRQ sources in this sample is presented, and the authors find the best fit for a power law Lorentz factor distribution to be $b=-3.33 \pm 1.30$ at $p=5$ or $b=-1.95\pm1.53$ for $p=7$. In this case the authors fit the redshift de-evolved x-ray luminosity function, which depends on the $\Gamma$ distribution due to beaming. They choose $\Gamma_{\text{min}}=5$ and $\Gamma_{\text{max}}=40$ from average properties of radio-loud blazars.

There are two best fit values for this sample, as the analysis is performed with different beaming parameters $p$. This parameter describes the beaming of a source depending on the emission mechanism and geometry of the emitting region.
\begin{equation}
    S_{\nu,\theta}=S'_{\nu}\delta^p
\end{equation}
where $\delta$ is the Doppler factor, $S$ the flux in the rest frame, $S'$ the observed frame flux and $p$ is the beaming parameter, dependent on spectral index $\alpha$. While we do not assume a flux beaming relationship for the BHXRBs, as we are only modelling jet kinematics and not luminosities, this is important for the AGN analysis as it is necessarily flux-limited. To calculate this parameter it is usually assumed that a jet can be approximated as a series of optically thick spheres. Depending on this geometry and the emission mechanism, $p$ is then related to the spectral index $\alpha$ in the observed frame \citep{dermer_beaming_1995}. For the self-Compton scenario, this relation would be $p=3 + \alpha$ \citep{sikora_comptonization_1994}, and for the external Compton process dominated emission the relation would be $p=4+2\alpha$ \citep{ghisellini_bulk_1989}. Both are plausible emission scenarios for the energy bands in which this sample is selected. The distribution of the spectral index value can be determined observationally, but the intrinsic distribution is not well understood. In the case of the BASS sample, the authors find that on average $\alpha=1.78$. Therefore, a parameter space in $p$ is explored in these population studies.

The fit for the BASS sample is conducted with a $\chi^2$ best fit and since multiple $p$ values give similar $\chi^2$ values, both are quoted. The authors state that higher values of $p$ are less likely, but a luminosity function turn-over at low luminosities would need to be detected to constrain this value further.

\subsubsection{\textit{Fermi}LAT Blazar sample}
Another sample for which this approach has been used, is selected by the \textit{Fermi} Large Area Telescope (LAT) in its first year of operation. The AGN are detected in the $10 \text{keV} -300 \text{GeV}$ range and flux-limited by a total flux in the $100 \text{MeV} -100\text{GeV}$ band of $F_{100} \approx 10^{-8}\text{photons~cm}^{-2}\text{s}^{-1}$ \citep{ajello_luminosity_2012}. Similarly to \cite{marcotulli_bass_2022} these authors also fit the de-evolved \textit{Fermi} luminosity function and find $b=-2.03 \pm0.70$ for $p=4$ and $b=-2.43\pm0.11$ for $p=5$. The $\chi^2$ fit is slightly worse for the $p=5$ case. The authors also use the same limits on $\Gamma$ as presented for the BASS sample. The authors test other Lorentz factor distributions, but the power law is statistically the best fit.

\subsubsection{AGN population study comparison}
\begin{figure*}
    \centering
    \includegraphics[width=0.8\textwidth]{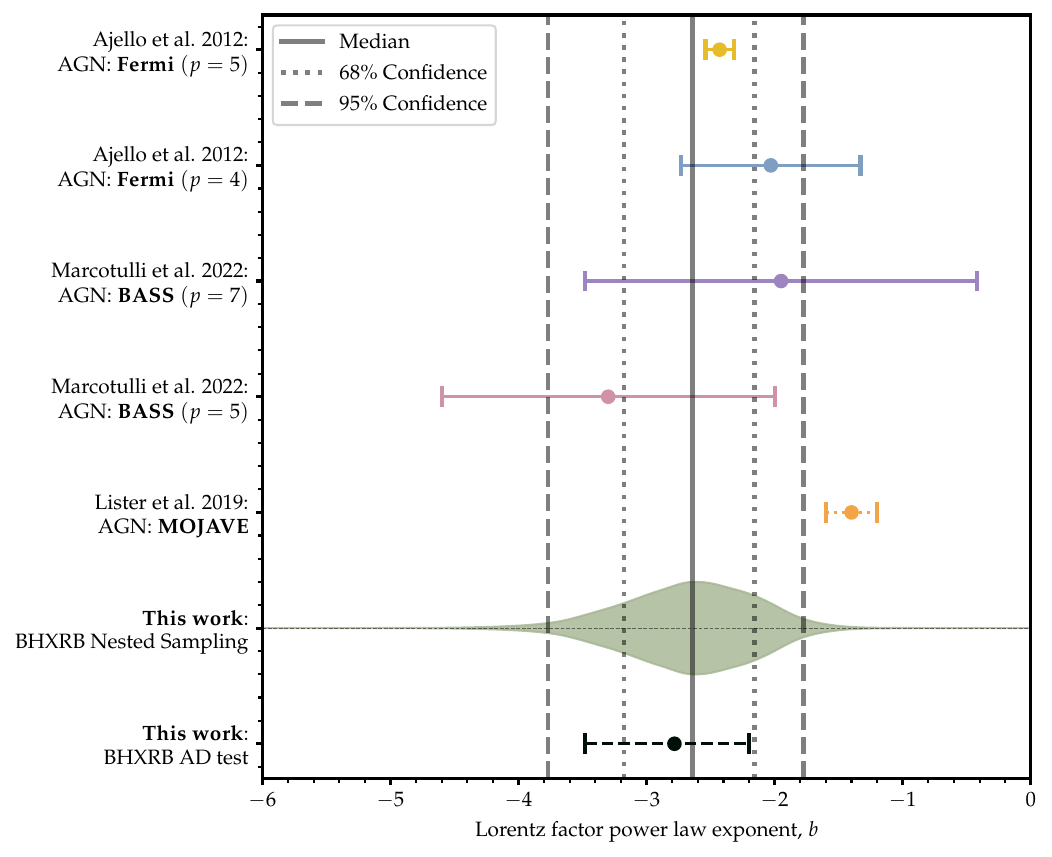}
    \caption{The comparison of AGN power law exponents with the BHXRB results in this work. The solid error bars come from $\chi^2$ model fitting, while the dotted error bars indicate that this error comes from Monte Carlo simulations and describe the bin width between different models. The dashed error bars for the BHXRB AD test case are described in Section \ref{sec:modelsel}.}
    \label{fig:b-exponents}
\end{figure*}

All AGN Lorentz factor distribution power law exponents are presented in Figure \ref{fig:b-exponents}, alongside the results for the BHXRBs found in this work.
The figures show that there is a spread within the values obtained for the various samples, but there is at least one result per survey that is consistent with all other samples. It needs to be taken into account that the results by \cite{lister_mojave_2019} obtain constraints on the $b$ parameter by fitting multiple observed distributions simultaneously. All other estimates of $b$ are based on only fitting a single distribution, for the BASS and \textit{Fermi} samples, this is the luminosity function. Additionally, there is a difference in the assumptions of the Lorentz factor limits.
 
There are also indications that there may be a population difference in sources selected in different wavelength bands.
\cite{lister_connection_2009} indicate that \textit{Swift}-BAT selected sources (BASS) tend to have faster estimated jet speeds than non-BAT selected AGN. \textit{Swift} light curves are much more sparsely sampled than \textit{Fermi}-LAT, which may also contribute to these effects \citep{hovatta_relativistic_2020}. This may then skew the analysis towards flatter power laws, because brighter and faster sources will occupy the tail of the power law and flatten the distribution. 
Additionally, since the \textit{Swift}-BAT and \textit{Fermi}-LAT selected sources do not have kinematic measurements, the Lorentz factor inferred for this population of sources likely traces a different region in the jet. This is because the flux at different wavelengths is emitted in different parts of the jet \citep{hovatta_relativistic_2020}. This possibly explains some of the discrepancy in inferred Lorentz factors for these differently selected samples.

In some AGN, and especially blazars, multiple ejections are observed, many with different apparent speeds. Only the fastest observed apparent speed is used in the population analysis by \cite{lister_mojave_2019}. The maximal flux is used in all population studies, which similarly corresponds to the most boosted jet component. This is likely the most accurate tracer of the maximal Lorentz factor of the sources due to the superluminal motion geometric effects discussed in Section \ref{sec:xrb_data}. Nevertheless, it can be argued that this selection does not fully capture the diversity of the jet population. By focusing exclusively on the highest speeds, the resulting power-law distribution of Lorentz factors may appear artificially flattened. 

In any case, the choice of using a singular $\Gamma$ to describe the jets of BHs is a simplifying assumption. Due to the high resolution achieved in studying SMBHs with VLBI observations, it has been revealed that there is a radial structure present in jets, which can be observed as "limb-brightening" of the jets \citep{kim_imaging_2024,tsunetoe_limb-brightened_2025}. Additionally, both stationary and moving components of the jets are detected, sometimes simultaneously \citep{weaver_kinematics_2022}. Multiple models have been proposed as explanations, such as spine-sheath models, where a fast-moving spine propagates within a slow-moving sheath \citep{ghisellini_structured_2005,chhotray_radiative_2017}. The interaction between the two components can lead to variable emission. In addition, there are shock-in-jet \citep{marscher_models_1985} and jet-in-jet models \citep{giannios_fast_2009} that aim to explain multi-wavelength emission variability and the moving blobs seen in VLBI observations by different components of the jet emitting over time due to disturbances and shocks. The debate about these models is fueled by the so-called Doppler factor crisis, which cannot reconcile the Lorentz factors derived from gamma-ray emission with those derived from kinematics \citep{bottcher_multi-wavelength_2019}.

In addition to population studies, there has also been work to derive the Lorentz factor parent distribution of blazars from the optical fundamental plane of BH activity \citep{saikia_lorentz_2016}. The results are consistent with the AGN population studies cited above.
\subsection{Comparing BHXRBs to AGN}
The BHXRB results are shown in Figure \ref{fig:b-exponents}. The vertical bars indicate the median and the $1 \text{~and~}2\sigma$ percentiles of the posterior. We also present the median and a probable range at $95\%$ confidence derived from the Anderson-Darling test as described in Section \ref{sec:adtest}. This range should not be interpreted as "real" uncertainties; it is an indication of confidence levels over the parameter space.

Nonetheless, the two populations inhabit a similar parameter space. The BHXRBs appear to follow a distribution with a lower exponent value than the AGN, except for one value from the analysis by \cite{marcotulli_bass_2022} using $p=5$. Most AGN exponent values lie at the high tail end of the BHXRB-derived distribution. Especially the result by \cite{lister_mojave_2019} is discrepant at $2\sigma$, but cannot be rejected at $3\sigma$ confidence. While this may indicate that BHXRBs have a steeper Lorentz factor distribution, it does need to be accounted for that the AGN analysis uses the maximum speeds for each source. For the BHXRBs we include the most representative speed for each source, except for MAXI $1820+070$, as mentioned in Section \ref{sec:xrb_data} we have reason to believe that these are fundamentally different jet types. GRS 1915+105 on the other hand has more than 10 ejections, which are consistently ejected at the same apparent speed (since 1997; see discussion in \citealt{miller-jones_deceleration_2006}). Therefore, we choose this mean value as representative for the source. In most other cases, only one ejection is observed \citep{fender_speeds_2025}. Other work by \cite{saikia_lorentz_2019} use the infrared excess to derive the Lorentz factor distribution of hard-state BHXRB jets. They suggest that the hard-state jets may follow a flatter power law.

When looking at different mass scales, we are also necessarily observing at very different time and distance scales when normalising in terms of $R_G = \frac{2GM}{c^2}$ and $t_G = \frac{2GM}{c^3}$ for the two populations. We observe AGN for a very short period of their duty cycle but we can resolve them much closer to the central supermassive black hole in units of gravitational radii. Conversely, BHXRBs are observed over multiple state changes. So, while they might be younger systems in absolute terms, we likely see them moving through their duty cycle at a much higher rate as compared to the AGN due to their smaller black hole masses. Nonetheless, we can only resolve BHXRB jets as separate ejecta at larger distances relative to their gravitational radii due to their much smaller size $\left(R \sim \mathcal{O}\left(10^{10}-10^{13}\right) R_{G} \right) $. Due to the discrepancy in relative distance scales, it would not be surprising to see differences in these two populations, yet this work has shown that there is intriguing agreement in the parent population $\Gamma$ distribution. 

Even if BHXRB jets were observed to be slower, one needs to consider that the jet in the mass-scaled AGN zone $\left(R \sim \mathcal{O}\left(10^4-10^6\right) R_G \right)$ much closer in to the BH, must have been the same speed or faster. Addtionally, BHXRB transient jets are considered discrete ejections. It is unlikely for these ejecta to have been significantly accelerated further away from the BH. In fact, Kelvin-Helmholtz instabilities and turbulence will necessarily slow down the jet as it propagates outwards \citep{bodo_deceleration_2003,savard_relativistic_2025}. There is evidence from the blazar population that indicates acceleration of the jet up to $100 \text{pc}$, while it is still collimating \citep{homan_mojave_2015}. A recent striped-jet model has been introduced to explain this variation \citep{giannios_grb_2019}. In units of gravitational radii $\left(R \sim \mathcal{O}\left(10^5-10^9\right) R_G \right)$ however, this distance is still well within the unresolved distance from the BH for BHXRBs. Due to the much better angular resolution for AGN jets we can see and model phenomena that we would simply not be able to resolve in XRBs even if they were there.

Multiple interesting attempts have been made in unifying BHs across the mass range and finding BHXRB accretion states in AGN \citep{kording_accretion_2006,svoboda_agn_2017, fernandez-ontiveros_x-ray_2021,moravec_radio_2022,kang_disk-jet_2025}. Within the class of blazars specifically, there have been suggestions that, going beyond the traditional classification of FSRQs and BL Lacs, there are two types of jets. Those with quasi-stationary knots and those with relativistic motion \citep{hervet_innovative_2016}. This could correspond to steady versus transient jets in BHXRBs. A distinction in these classes has also been found in the variation of the core position across wavelength \citep{hovatta_relativistic_2020}. The FSRQs which are used in the parent population analysis of AGN correspond to the second class of jets and have been likened to BHXRBs in their transition from hard-to-soft state \citep{kang_disk-jet_2025}. Therefore, it would be reasonable to conclude that the comparison presented here investigates the same type of large-scale transient jet across the mass range.

We have demonstrated that the apparent striking difference in observed $\beta_{\rm app}$ in BHXRBs and AGN can be largely due to inclination and beaming effects and may not be due to an intrinsic difference in Lorentz factors. Instead, it is impossible to constrain maximum Lorentz factors based on kinematics alone and it is likely that the Lorentz factors follow a power law distribution, which is steep, but allows for large Lorentz factors in a small fraction of the sample.
The size of the parent population needed to produce the observed sample can give insight into expected observations of blazar-like XRBs. \cite{lister_mojave_2019} find that the parent population of AGN to produce a flux-limited sample of 174 sources consists of around $3\times10^6$ sources. Therefore, we are only seeing around $0.1\%$ of the parent population and only the most extreme objects. There are many less relativistic and more inclined sources in the universe that do not make it into a flux-limited sample. As the BHXRB sample only contains $19$ sources, there is a very small probability for one of the sources to be highly aligned with the line-of-sight only given the isotropic inclination distribution. Adding the small probability of high Lorentz factors due to the power law parent distribution, this makes it obvious that a much larger BHXRB sample is required to observe blazar-like BHXRB sources. In fact, a source flaring similar to MAXI J1820+070 as presented in \cite{bright_extremely_2020} could be seen up to $21.2$ Mpc at a detection limit of $S_{\rm lim} = 0.1$ mJy if it had $\Gamma = 10$ and $\theta=3^\circ$ (approximate mean values of MOJAVE AGN sample). Given the best-fit parent population distributions, sources with $\Gamma \geq 10$ and $\theta \leq 3$ make up only $0.0028\%$ of the parent population. Although the detectable volume would include the Virgo cluster, it is likely that given the duty cycles and typical flare durations of BHXRBs ($6.7$ hrs in the case of MAXI J1820+070) these objects would be missed even when flaring brightly given the observing cadences of current radio surveys of these galaxies.

\subsection{Theoretical interpretation}
A very common interpretation of the moving components in AGN jets is the commonly discussed shock-in-jet model \citep{marscher_models_1985}, which is a promising framework to explain jet variability and emission. This model aims to produce observable jet characteristics with a shock moving through a steady jet and has been successful in explaining flares from AGN and the moving blobs observed within blazar jets especially when taking into account emission mechanisms and boosting \citep{fromm_spectral_2016,fichet_de_clairfontaine_flare_2022}. Observations have supported the idea of an underlying steady jet by resolving stationary features in blazar jets, which could be standing shocks \citep{agudo_jet_2001,hervet_shocks_2017,weaver_kinematics_2022}.
Similarly, it has been found that the observed ``moving blobs'' are usually linked to radio flares \citep{lindfors_synchrotron_2006}, which has been successfully reproduced in simulations of moving shocks through jets \citep{fromm_spectral_2016,fichet_de_clairfontaine_flare_2022}. \citep{clairfontaine_flux_2021} also show that the shocks become energised when moving through the rarefactions of the standing shocks, which causes a strong variability in their Lorentz factor. It is possible that the superposition of many such processes causes the observed power law distribution.

The shock-in-jet model is not only successful in explaining AGN behaviour, but has also been applied to XRBs. 
Especially the rapid variability and the spectral shape of the ``hard state'' in XRBs can be successfully explained by this model \citep{turler_shock--jet_2010,malzac_spectral_2014}.
Another important mean of analysing the variability in both XRBs and AGN is the power spectral density (PSD). In SMBHs the PSDs can be fitted with power laws \citep{kankkunen_active_2025}, whose slopes are associated with different noise processes. The PSD slopes in different wavelength bands vary in blazars \citep{hovatta_relativistic_2020}. In some XRBs PSD break frequencies can be found with sufficient light curve sampling. These likely correspond to characteristic time scales of the system, such as the inner disk edge or cooling timescales \citep{zdziarski_radiative_2004,ishibashi_physical_2012}. \cite{mchardy_active_2006} also show that there is a scaling of break frequency with BH mass, which suggests that there is a similarity in the innermost accretion processes across the BH mass range.
\cite{malzac_emission_2012} and \cite{jamil_ishocks_2010} use X-ray PSD slopes to inject Lorentz factor variability into the jet base, which is successful in producing observed spectra for hard state XRB jets. This association of variability and observables in the jet could be used to explain the power law Lorentz factor distribution found in this work. Future work will investigate this connection further.
Despite its success in reproducing observables, the shock-in-jet model has not remained unchallenged and the jet-in-jet model has been proposed instead. A combination of models, also incorporating the sheath-spine structure has also been tested \citep{macdonald_through_2015} and is probably a likely explanation for the observed phenomena. 

One question to consider is why the distribution of Lorentz factors is so well described by a power law, preferentially over any other distribution, such as Gaussian, exponential or flat. Indeed, \cite{smith_distribution_2000} find that the velocity distribution of shocks from driven supersonic turbulence tends to a power law distribution proportional to the square root of the shock speed. The authors attribute the specific exponent to the Gaussian driving term. The power law tail is caused by the superposition of driving waves. This shows that many physical processes can lead to power law velocity distributions of shocks. If we considered the observed speeds in jets to come from moving shocks, as in the shock-in-jet model, the finding by \cite{smith_distribution_2000} could offer a physical explanation why the moving shocks would follow a power law distribution. The driving term within astrophysical jets could likely be caused by the fast variability, which can be observed from all of these sources \citep{margon_rapid_1984,merloni_fundamental_2003, vaughan_rapid_2005, homan_high-_2005,fender_eight_2007,bottcher_multi-wavelength_2019}. Similarly to the work by \cite{malzac_emission_2012} and \cite{jamil_ishocks_2010} the PSD describing the variability could inject shocks into the jets which then tend to a power law velocity distribution. This is an explanation of why a single jet would display a power law shock velocity distribution, superposed each of these could lead to a population-wide power law distribution. 

Once again we need to consider the difference in observed relative scales in the populations and its consequence on this analysis:
For BHXRBs the power law Lorentz factor distribution definitely presents a population-wide statistic as certain individual sources show the same ejection speed across years and decades, such as GRS 1915+105 \citep{miller-jones_deceleration_2006}. Since we are observing much larger relative distance scales in BHXRBs the observed distribution is possibly a superposition of the individual processes described in AGN. Whether we would see more variability and a similar structure to blazar jets on smaller scales is unknown. If BHXRBs possess a similar spine-sheath structure to blazars it would be likely that the observed emission stems from the sheath as the spine emission would be boosted out of the line-of-sight (e.g. \citealt{hardee_grmhdrmhd_2007,mizuno_three-dimensional_2007,qian_jet_2018,dihingia_thin_2024}). Blazar spines on the other hand become much brighter at small angles to the line of sight due to Doppler boosting. Work by \cite{spada_internal_2001} on internal shocks in blazars show that the internal shock scenario leads to an asymptotic bulk Lorentz factor at large radii as the shocks travel down the jet and interact. If this model could be applied to BHXRBs, it could explain the consistency of similar speeds of the ejected blobs in GRS 1915+105, for example.
Therefore, a blazar-like structure in BHXRB jets could be a promising way to unify the trends we are seeing in observables across both populations.
To further account for inclination effects and resolution differences in AGN and BHXRBs simulations could be a promising tool for further work. 

There is convincing evidence that shows that two types of jets contribute to the BHXRB parent population. It has been known for a long time that there are two types of jets associated to the two distinct spectral types of BHXRBs. In addition, new evidence by \cite{fender_speeds_2025} suggests that there can be an additional distinction in jet population among the large-scale, transient jets: one which is spin-locked and fast and one disc-driven population, which is slower. These two distinct types of jets are even observed within the same spectral state in the same source and therefore likely not an intrinsic feature of the source, but a transient phenomenon or another accretion "state" the source cycles through. This indicates that there could be a contribution of two components to the observed Lorentz factor distribution, with a larger contribution from the slow jets around the lower Lorentz factors and a fast tail of the spin-locked jets. Ultimately, to constrain this further, it is necessary to obtain a larger sample of observations. 

\section{Summary and Conclusions}
This work presents a novel analysis of BHXRB jets and their apparent speeds. For the first time, it is possible to perform parent population analysis of these jets. First we explore the parameter space using the traditional AD test method. We are also able to introduce nested sampling with a novel likelihood definition to this problem. Through consistency with the AD test results, we show that it is possible to define a meaningful likelihood to compare the observed apparent speed distribution to the model distribution. This enables us to find full Bayesian posteriors for all parameters and additionally use Bayesian evidence to compare various models for the Lorentz factor distribution.

We find that BHXRB jets are best described by a power law $\Gamma$ distribution, similar to AGN. We find a best fit value of $b = -2.64_{-0.55}^{+0.46}$ for the power law exponent. This is consistent with both a \textit{Fermi}-LAT and a \textit{Swift}-BAT selected sample within $1 \sigma$. The radio-selected MOJAVE sample is discrepant at $2\sigma$, which may hint at a fundamental difference between the populations. There are, however, many fundamental challenges to the comparison to be considered, such that we are looking at very different distance scales in $R_G$ of the jets and acceleration may play an important role, especially in the AGN. Different wavelengths probe different parts of the jets, which additionally complicates direct comparison. Additionally only the maximum observed speed and flux are used in the AGN population studies. 

Despite all of these complications, strikingly, both the BHXRBs and the AGN occupy a similar parameter space in the power law exponents.
The difference in apparent speed distribution between these populations stems largely, possibly entirely, from inclination and beaming effects. The agreement is especially remarkable due to the difference in mass and distance scales between these populations. It is also obvious that the AGN population is dominated by the most extreme sources. Due to its flux-limited nature and the extreme boosting effects the parent population of this sample is very large. This is not the case for the XRBs. Due to the X-ray selection of the sample, they are subject to very different selection effects. As we were able to determine the shape of the parent population, this enables us to show that due to the limited sample size there is a very low probability to find highly alligned and fast sources that are like blazars, as these make up only a very small percentage of the parent population.

The upper limit of the Lorentz factor distribution of BHXRBs is completely unconstrained. This is the result of geometric effects on superluminal motion. Therefore, based solely on kinematics, BHXRBs could have very high Lorentz factors, similar to AGN. Other non-kinematic arguments may be able to constrain this upper Lorentz factor limit further.

With a growing XRB sample this analysis will continue to become more constraining. Additional work simulating jets to constrain energy estimates better, will be extremely beneficial in further refining our understanding of BH jets at all mass scales. Additionally, it would be interesting to account for the variability in blazar jets better to investigate whether this changes our understanding of the underlying distributions.

\section*{Acknowledgements}

The authors acknowledge valuable conversations with Alan Marscher and Matthew L. Lister.

The authors thank the anonymous referee for their insightful comments and constructive report.

CL acknowledges support from the ERC "Blackholistic" grant.

RPF acknowledges support from UKRI, The ERC and The Hintze Family Charitable Foundation. 

JM acknowledges funding from a Royal Society University Research Fellowship (URF$\backslash$R1$\backslash$221062).

We gratefully acknowledge the use of the following software packages: \texttt{matplotlib} \citep{Hunter:2007}, \texttt{scipy} \citep{2020SciPy-NMeth}, \texttt{dynesty} \citep{speagle_dynesty_2020,koposov_joshspeagledynesty_2024}.
\section*{Data Availability}

The BHXRB data can be accessed as part of the data published by \cite{fender_speeds_2025}. And the AGN data used to plot can be found in \cite{lister_mojave_2019} and \cite{homan_mojave_2021} where the data is available in machine-readable format. 



\bibliographystyle{mnras}
\bibliography{20251107-2_bib} 


\appendix

\section{Alternative Lorentz factor distributions}
\label{app:altLorentz}
The alternative models tested for the parent population Lorentz factor distribution are:

Truncated Gaussian Distribution:
$$N(\Gamma) = \frac{\frac{1}{\sigma\sqrt{2\pi}} e^{-\frac{1}{2}\left(\frac{\Gamma-\mu}{\sigma}\right)^2}}{\Phi\left(\frac{\Gamma_{\text{max}}-\mu}{\sigma}\right) - \Phi\left(\frac{\Gamma_{\text{min}}-\mu}{\sigma}\right)} \quad \text{for } \Gamma_{\text{min}} < \Gamma < \Gamma_{\text{max}}$$
Where $\Phi$ is the cumulative distribution function of the standard normal distribution.

Truncated Exponential Distribution:

$$N(\Gamma) = \frac{\lambda e^{-\lambda \Gamma + \lambda\Gamma_{\text{min}}}}{1 - e^{-\lambda(\Gamma_{\text{max}}-\Gamma_{\text{min}})}} \quad \text{for } \Gamma_{\text{min}} < \Gamma < \Gamma_{\text{max}}$$

Truncated Gamma Distribution:

$$N(\Gamma) = \frac{\beta^\alpha \Gamma^{\alpha-1} e^{-\beta \Gamma}}{\gamma(\alpha,\beta\Gamma_{\text{max}}) - \gamma(\alpha,\beta\Gamma_{\text{min}})} \quad \text{for } \Gamma_{\text{min}} < \Gamma < \Gamma_{\text{max}}$$

Where $\gamma(\alpha,x)$ is not the Lorentz factor, but the lower incomplete gamma function defined as $\gamma(\alpha,x) = \int_{0}^{x} t^{\alpha-1} e^{-t} dt$.

\begin{figure}
    \centering
    \includegraphics[width=1\linewidth]{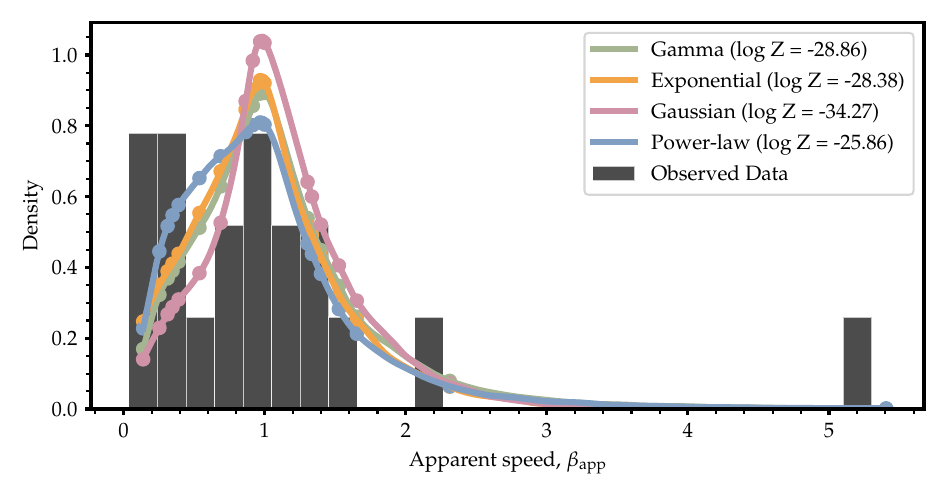}
    \caption{The histogram shows the observed apparent speeds of the BHXRB sample. Overplotted are the KDEs of the best-fit models for each of the different underlying Lorentz factor distributions. The dots indicate the observed apparent speeds at which the KDEs are evaluated to obtain the likelihood for the nested sampling.}
    \label{fig:ap_model_comp}
\end{figure}


\bsp	
\label{lastpage}
\end{document}